\documentclass[nanomaterials,article,submit,moreauthors,pdftex]{mdpi} 

\preto{\abstractkeywords}{\nolinenumbers}

\firstpage{1} 
\pubvolume{1}
\issuenum{1}
\articlenumber{0}
\pubyear{2021}
\copyrightyear{2020}
\datereceived{} 
\dateaccepted{} 
\datepublished{} 
\hreflink{https://doi.org/} 

\Title{Anisotropic Optical Properties of Hexagonal Boron Nitride Thin Films}

\TitleCitation{Anisotropic optical properties of hexagonal boron nitride thin films}

\Author{L.V. Kotova $^{1,*}$\orcidL{}, L.A. Altynbaev $^{1}$, M.O. Zhukova $^{2}$\orcidM{}, B.T. Hogan $^{3,4}$\orcidA{}, A. Baldycheva $^{3}$\orcidB{} and V.P. Kochereshko $^{1}$}

\AuthorNames{Lyubov Kotova, Linar Altynbaev, Maria Zhukova , Benjamin Thomas Hogan, Anna Baldycheva, Vladimir Kochereshko}

\AuthorCitation{Kotova, L.; Altynbaev, L.; Zhukova, M.; Hogan, B.T.; Baldycheva, A.; Kochereshko, V.;}

\address{%
$^{1}$ \quad Ioffe Institute; 194021 St. Petersburg, Russia\\
$^{2}$ \quad ITMO University; 197101 St. Petersburg, Russia\\
$^{3}$ \quad Department of Engineering, University of Exeter; Exeter EX4 4QF, UK\\
$^{4}$ \quad University of Oulu; 90014 Oulu, Finland}

\corres{Correspondence: kotova@mail.ioffe.ru}

\abstract{
Few-layer flakes of hexagonal boron nitride were prepared by ultrasonication of bulk crystals and agglomerated to form thin films. The transmission and reflection spectra of the thin films were measured. The spectral dependences of the linear and circular polarization revealed a hidden anisotropy of the films over the whole sample area which could not be explained by the anisotropy of the chaotically-oriented individual particles. Statistical analysis of optical microscopy images showed a macroscopic particle density distribution with ordering corresponding to the optical axis observed in the polarization data.  
}


\begin{document}

\section{Introduction}

Interest in atomically thin (2D) materials has increased exponentially since the unique properties  of graphene were first reported. Since then, researchers have extended the family of known 2D materials to include MXenes, transition metal dichalcogenides, and boron nitride amongst others. The diverse properties of 2D materials combined with their ultrathin nature open up the prospect of creating van der Waals heterostructures, in which different two-dimensional materials are stacked layerwise \cite{Geim}. Due to the contrasting strong intralayer covalent bonding and relatively weak interlayer interactions, van der Waals heterostructures retain many properties of the individual layer materials. Therefore, to fully unlock the potential of van der Waals heterostructures requires a full understanding of the properties of the individual 2D material layers. 

2D materials can generally be synthesised by two overarching routes: top-down and bottom-up. Bottom up methods (such as chemical vapour deposition or molecular beam epitaxy) can produce high quality 2D materials over relatively large areas, but are time-consuming, energy-intensive, and cost-prohibitive. Top-down methods involve exfoliation of the bulk crystal. The simplest and most common technology for exfoliating bulk crystals is mechanical cleavage of the layers using sticky tape. This method can produce monolayer single crystals with very few defects. However, the size of the individual monolayer flakes is generally small and the yield low, such that the method cannot be easily scaled to mass production. New methods are hence being developed to overcome these respective fabrication barriers.

One of the most promising developing technologies devoid of these shortcomings is the cleavage of thin layers from a bulk crystal using ultrasonication in a liquid. This method can be easily applied to bulk crystals where the interlayer interactions are weaker than the intralayer. The bulk crystals are dispersed in a solvent through which ultrasonic waves are propagated. The energy of the ultrasonic waves causes cavitation to occur within the bulk crystal, resulting in the delamination as the layers are pushed apart. The result is a solution containing flakes of different sizes and thicknesses which can then be further separated by centrifugation and fractionation of the solution. 
Bulk hexagonal boron nitride (\textit{h}-BN) has a similar layered structure to graphite which can be readily exfoliated to give thinner crystals down to the monolayer limit. Since the interlayer interactions in \textit{h}-BN are weak compared to the intralayer covalent bonding, cleavage occurs relatively easily. Monolayer (2D) \textit{h}-BN is stable with a graphene-like lattice structure consisting of alternating boron and nitrogen atoms (see Figure~\ref{fig:BN}~a). \textit{h}-BN is an attractive material for barrier layers in heterostructures as it is one of the strongest known electrically-insulating materials, with significant prospects for use in nanoelectronics \cite{Moon} and optoelectronics \cite{Ronning}. 

In this paper, thin films of \textit{h}-BN flakes were produced by exfoliating bulk \textit{h}-BN crystals \textit{via} ultrasonication. Optical characterization of these samples was carried out. In particular, we studied the spectral dependences of the Stokes parameters of light passed through the sample under normal and oblique incidence in linear and circular polarizations.

\section{Experiment } 
\begin{figure}[h]
 \centering
  \includegraphics[width=1\linewidth]{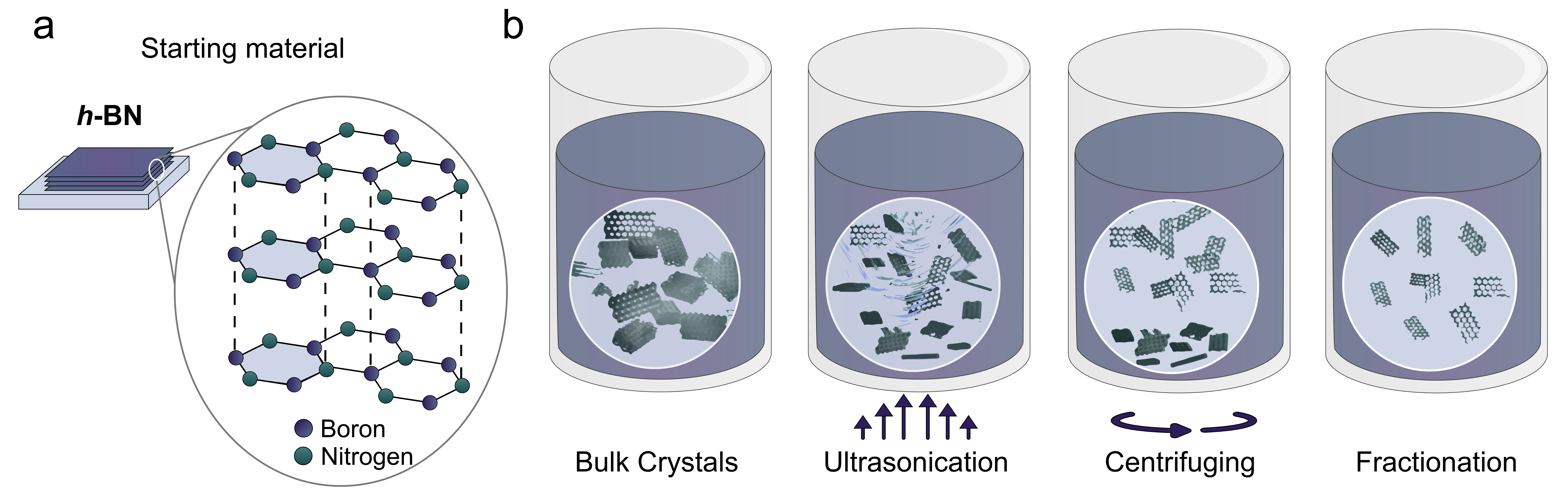}
  \caption{(a) The \textit{h}-BN structure. (b) The \textit{h}-BN exfoliation procedure:  bulk \textit{h}-BN powder is dispersed in a mixture of isopropanol and deionized water, the solution is ultrasonicated, then centrifuged, and finally the supernatant is extracted.}
  \label{fig:BN}
\end{figure}

Hexagonal boron nitride thin films were prepared on polyethylene terephthalate (PET) substrates by the following method (see Figure~\ref{fig:BN}~b for an illustration). Bulk crystal \textit{h}-BN powder (25~g) was initially dispersed in a beaker  containing 500~mL of a 60:40 ratio mixture of isopropanol and deionized water. The dispersion was ultrasonicated in a 120~W ultrasonic bath filled with deionized water at an initial temperature of 298~K. Five hour-long periods of ultrasonication, separated by 30 minutes each to prevent excessive heating of the solvent (over each one hour run, the temperature of the bath increased to around 318~K, which returned to around 298~K after the period of inactivity), to ensure a sufficient exfoliation yield from the bulk crystals. The partially exfoliated dispersion was then put through a process of centrifuging for 10 minutes at 2000~rpm to sediment out the residual bulk material and larger particles, narrowing the distribution of particle sizes present in the supernatant and favouring thinner particles. The dispersion was fractioned to extract the supernatant, ensuring only suitably sized particles remained. The supernatant was then dried under vacuum (~0.1~atm, 343~K) in a Schlenk line to fully remove the solvent mixture. 606~mg of the resulting powder was then redispersed in 12~mL of isopropanol mixed with 8~mL of deionized water to give a solution with a known concentration of around 30~mg/mL. This redispersed solution was ultrasonicated for a further 30 minutes to ensure a uniform dispersion.

The redispersed solution was filtered through a microporous ($\sim$0.2 $\mu$m pores) PTFE (polytetrafluoroethylene) membrane to produce homogeneous thin films from the exfoliated \textit{h}-BN. High quality films have been successfully developed using other two-dimensional materials (for example from dispersions of graphene oxide and molybdenum disulfide \cite{Hogan,Akbari,Fu}), but not previously for \textit{h}-BN. The thin films produced from filtering of the dispersion was achieved in accordance with a previously described method \cite{Jalili, Shaban}, marking a facile and scaleable method towards large-scale integration of the two-dimensional material. The membrane holding the thin film was carefully transferred to the desired (PET) substrate, with the side with the filtered \textit{h}-BN in contact with the substrate surface. Isopropanol was used to wet the membrane. The substrate was then heated to 343~K to evaporate the isopropanol and simultaneously release the \textit{h}-BN film from the membrane. The membrane was then removed leaving the thin film supported on the substrate.

\begin{figure}[h]
\centering
  \includegraphics[width=0.8\linewidth]{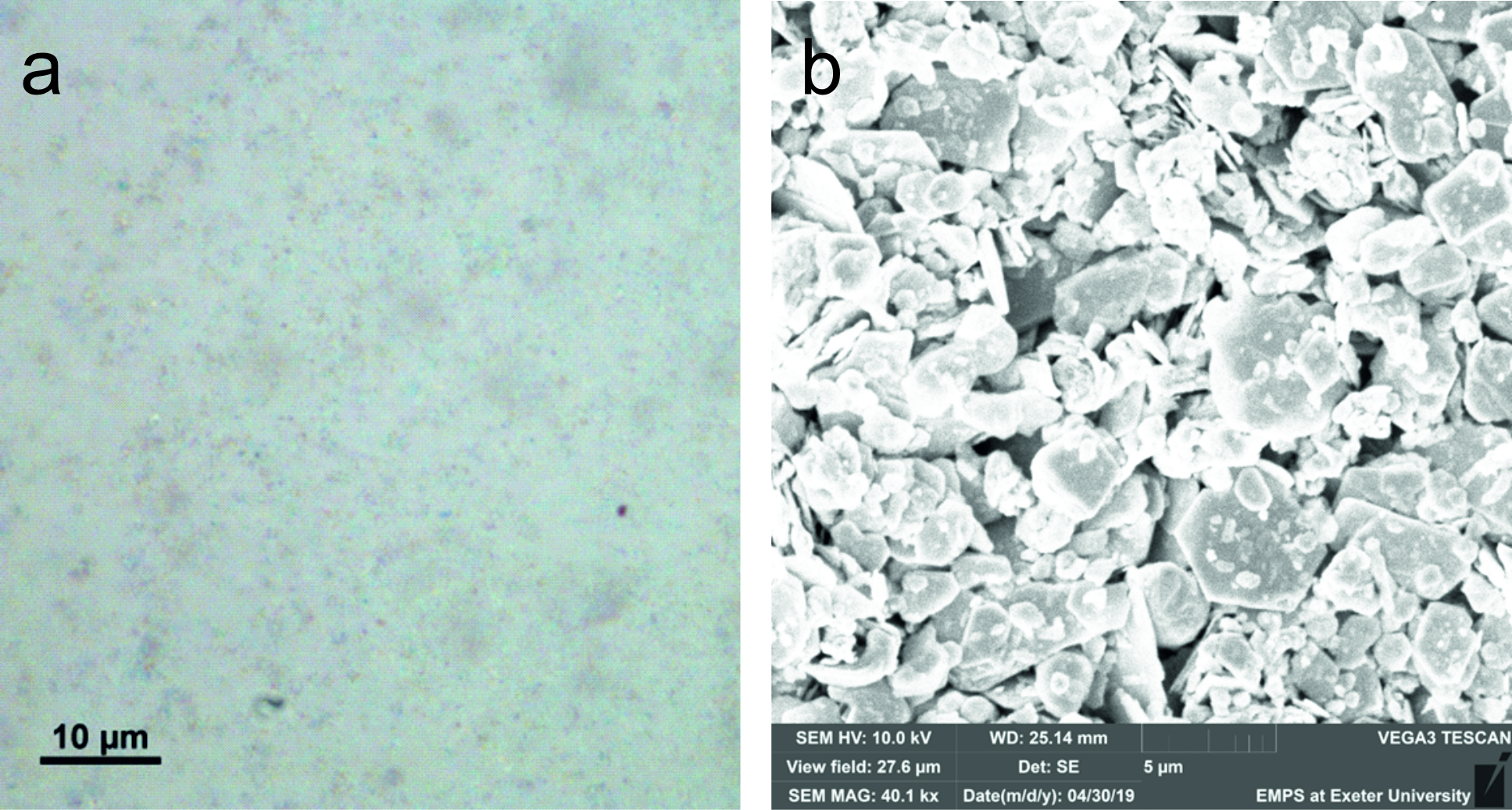}
  \caption{Images of the sample surface taken with: (a) an optical microscope at x100 magnification, (b) a scanning electron microscope with at x40000 magnification. Scale bars are shown in the images.
}
  \label{fig:SEM}
\end{figure}

The average overall thickness of the obtained films (formed of stacked individual thin \textit{h}-BN particles) was about 5 microns. Optical and scanning electron microscopy (SEM) images of the surfaces of the obtained samples are presented in Figures~\ref{fig:SEM}~a and b. The sample consists of thin crystalline plates of \textit{h}-BN with transverse sizes of up to 5~microns and thicknesses of 1-10 monoatomic layers. Visually, individual crystals appear to be chaotically distributed. There are no clear selected directions on the sample surface at either the hundred-micron scale or the micron scale. 

The Raman spectrum of the thin film showed that the crystal structure corresponds to that of perfect \textit{h}-BN crystals. At the same time, intense background scattering was observed, indicating a high density of defects in the overall film.

\begin{figure}[h]
 \centering
  \includegraphics[width=0.85\linewidth]{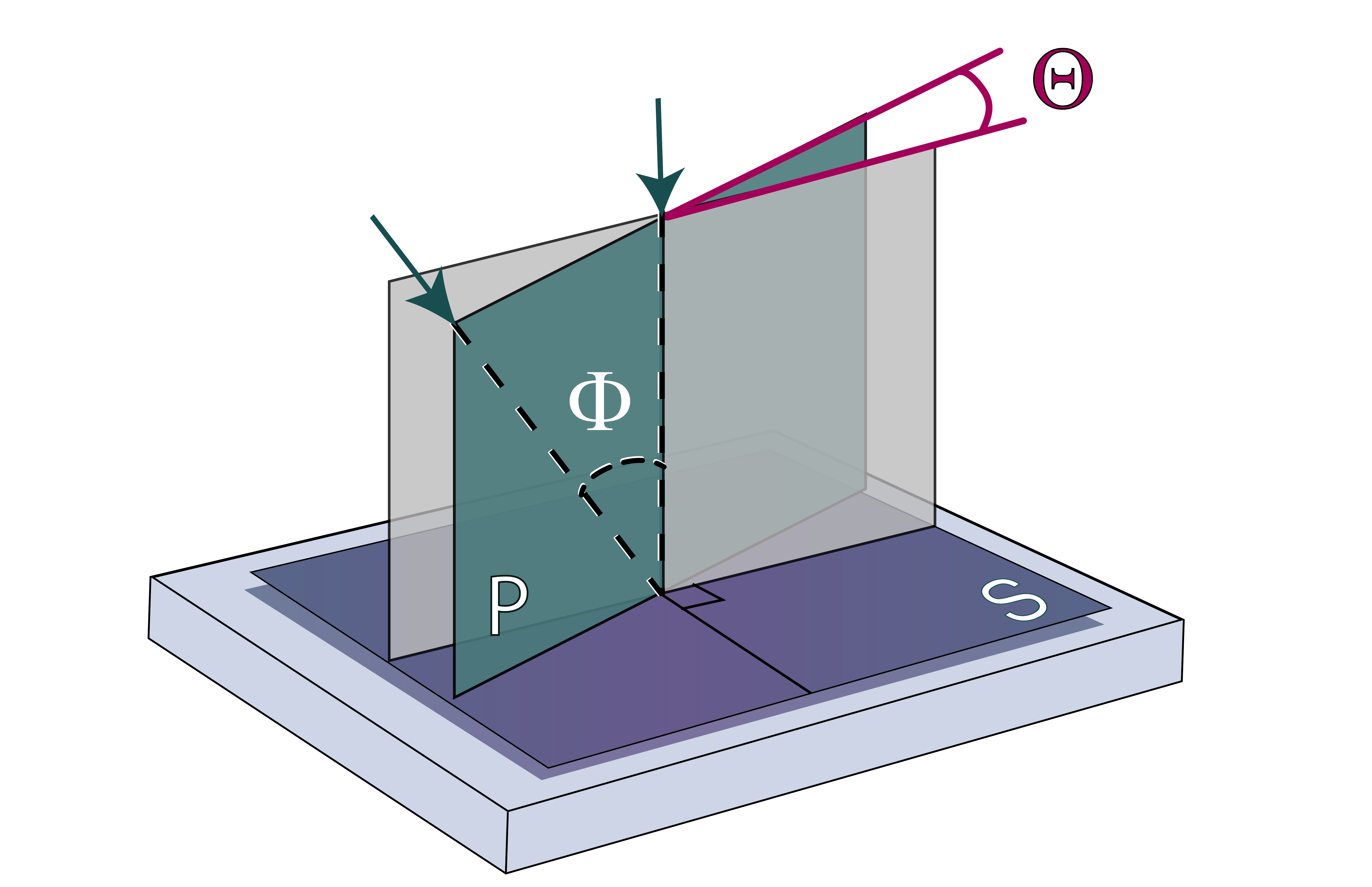}
  \caption{The experimental scheme. {\em P} is the plane of incidence, {\em S} is the sample surface, $\Phi$ is the angle of incidence, and $\Theta$ is the azimuthal angle between the plane of incidence and the edge of the sample.  }
  \label{fig:Setup}
\end{figure}

In Figure~\ref{fig:SEM}~a it can be seen that the sample appears matte, as strong Rayleigh light scattering was observed from the surface. We measured the spectra of Rayleigh scattering for linearly polarized light as a function of the incident angle ($\Phi$, see Figure~\ref{fig:Setup}) in the range from 0 to $\pm 45^\circ$ and the azimuthal angle ($\Theta$, Figure~\ref{fig:Setup}) in the range from $\pm 0-360^\circ$. The spectra did not have any structure, except for one broad maximum of 1~eV width at energy 1.7~eV. The intensity of the scattered light did not depend on the azimuthal angle $\Theta$. In the dependence on the incident angle $\Phi$, there were no qualitative differences between the spectra, except for a change in the total intensity. 
 
The polarized light transmission spectra of the samples were recorded at room temperature using a 0.5~M spectrometer equipped with a CCD~detector. A halogen lamp was used as the light source. There was a strong absorption in the region within the \textit{h}-BN bandgap at energies above 3~eV. No interference features were observed in the transparency range from $\sim$1.3 eV up to $\sim$3.0 eV, indicating that the sample thickness noticeably fluctuated at the scale of the light beam size ($\sim$1 mm). 

The Stokes parameters $P_{\rm circ}$ and $P_{\rm lin}$ of the transmitted light in the transparency range of this sample were measured for linearly polarized incident light. The degree of circular polarization $P_{\rm circ}$ is 

\begin{equation}
\label{S}
			P_{\rm circ}=\frac{I_{\sigma_+}-I_{\sigma_-}}{I_{\sigma_+}+I_{\sigma_-}},
\end{equation}
where $I_{\sigma_+}$ is the intensity of the transmitted light with right circular polarization, and $I_{\sigma_-}$ is the intensity of the transmitted light with left circular polarization. The degree of linear polarization $P_{\rm lin}$ is 

\begin{equation}
\label{L}
		P_{\rm lin}=\frac{I_{0}-I_{90}}{I_{0}+I_{90}}	,
\end{equation}

where $I_{0}$ is the intensity of the transmitted light with a linear polarization state that coincides with that of the incident light, and $I_{90}$ is the intensity of the transmitted light with a linear polarization state orthogonal to that of the incident light.

\begin{figure}[h]
 \centering
  \includegraphics[width=0.9\linewidth]{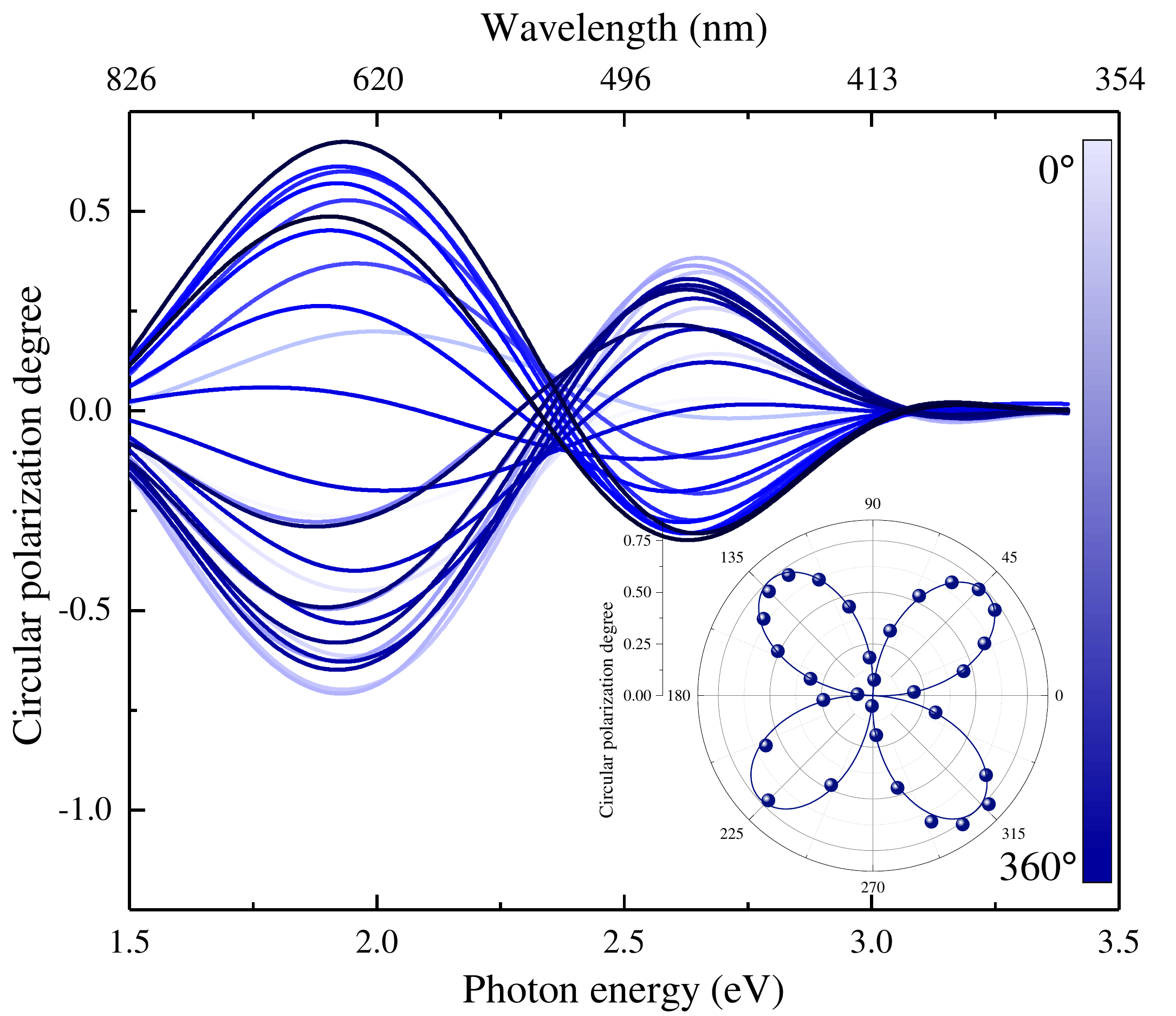}
  \caption{Spectral dependences of the circular polarization degree $P_{\rm circ}$ at azimuthal angles from 0 to 360 degrees for the light passed through the sample when the incident light was linearly polarized at room temperature. The insert shows angular dependence of the polarization conversion. }
  \label{fig:CircPol}
\end{figure}

Figure~\ref{fig:CircPol} shows the spectral dependencies of the degree of circular polarization $P_{\rm circ}$ as a function of the azimuthal angle $\Theta$ between the polarization direction and the sample edge at normal light incidence. When linearly polarized light passed through the sample, it acquired a circularly polarized component. The degree of circular polarization was dependent on the azimuthal angle $\Theta$. At an energy of 2.36~eV, the degree of circular polarization changed sign. At this energy, the sample is optically isotropic. The maximum value of circular polarization occurred at an energy of 1.91~eV.

\begin{figure}[h]
 \centering
  \includegraphics[width=0.9\linewidth]{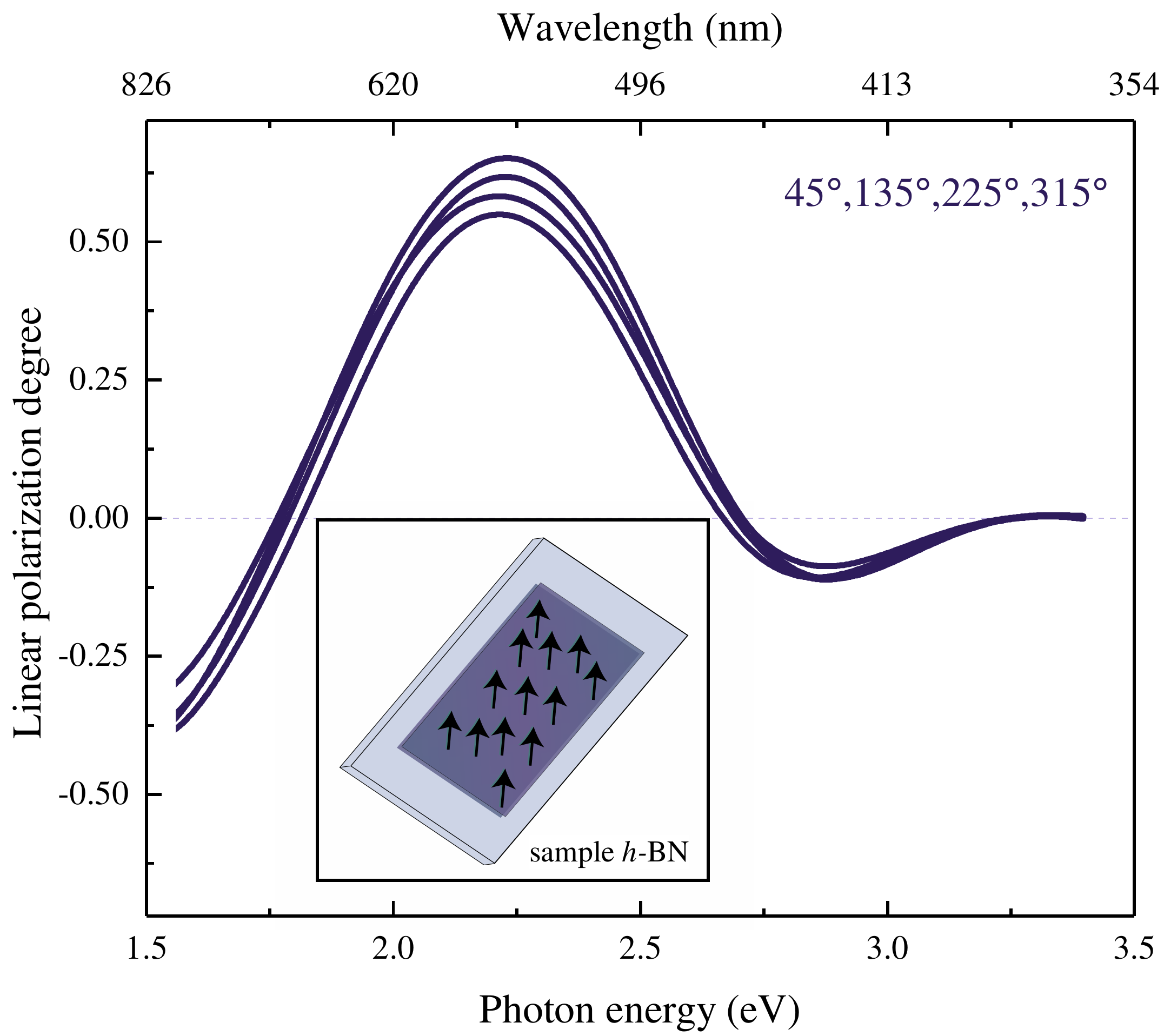}
  \caption{Spectral dependences of the linear polarization degree $P_{\rm lin}$ at selected azimuthal angles  for linearly polarized light passed through the sample. T=300 K. The inset shows the direction of the optical axis over the entire area of the sample. }
  \label{fig:LinPol}
\end{figure}

Figure~\ref{fig:LinPol} shows the spectral dependence of the degree of linear polarization $P_{\rm lin}$ in the axes coinciding with the direction of polarization of the incident light as a function of the azimuthal angle $\Theta$. For linear polarization, similar dependences to those for the circular polarization were obtained with one difference: at energies where $P_{\rm circ}$ was maximal, $P_{\rm lin}$ was minimal. The magnitude of the degree of linear polarization reaches a maximum when the plane of polarization coincides with the optical axis of the sample. 

The obtained data demonstrates that the birefringence phenomenon is observed in the entire measured spectral range. This phenomenon manifests itself in the conversion of linear polarization into circular polarization. 

The birefringence was measured at different points on the sample surface. It was found that the direction of the optical axis did not change over the entire sample area ($\sim$1 cm$^2$). The arrows show the direction of the optical axis in the inset of Figure~\ref{fig:LinPol}.


\section{Results and Discussion}

The appearance of an optical axis in the polycrystalline sample could be caused by one of two reasons: by anisotropy of the individual particles themselves or by anisotropy of their collective distribution \cite{Jacucci}. Individual crystals of \textit{h}-BN do indeed have optical anisotropy. The refractive indices of light along the hexagonal axis and perpendicular to it differ significantly \cite{Segure}. However, as shown in Figure~\ref{fig:SEM}, the orientation of microcrystals in our sample is completely chaotic (at least on scales from tens to hundreds of microns). Given the chaotic arrangement of the indiviudal particles in the thin film, the anisotropy of the particles alone cannot lead to an overall anisotropy of the entire sample. However, it could manifest as a local anisotropy. Thus, the observed birefringence with the same direction of the optical axis over the entire sample area ($\sim$1 cm$^2$) cannot be caused by the anisotropy of the \textit{h}-BN microcrystals themselves. This indicates that the cause of anisotropy is not hidden in the properties of individual microcrystals but is common to the entire sample and manifests itself at large distances. It was verified that the substrate itself is optically isotropic. Measurements of the reflected/scattered light from the sample showed that, in contrast to the case described in \cite{Kotova}, the inclination of microcrystals in our powder is chaotic. The presence of hidden order in a disordered substance may indicate that there is a small system of microcrystals in which there is no complete averaging of properties. However, this is not valid for our case since the anisotropy is observed in a large volume. Another reason may be the ordered arrangement of microcrystals, which manifests as optical anisotropy. A similar case was theoretically modeled in \cite{Jacucci} where no anisotropy was visible “by eye", but anisotropic light scattering occurred.

To describe the observed phenomenon, we can use the model of an effective dielectric medium \cite{Thouless,Landauer,Choy}. Although the particle sizes in our case are comparable or larger than the wavelength of light, and the applicability of this model is not entirely justified, the basic properties of the medium should be described at least qualitatively by this model.

We will consider our sample as a composite with an effective permittivity $\varepsilon_{eff}$ consisting of microcrystals and voids between them. We take the dielectric permittivity of boron nitride averaged over all orientations as an approximation of the dielectric permittivity of the ensemble of microcrystals $\langle\varepsilon\rangle = {\langle n \rangle}^2$  \cite{Segure}. We then use the Maxwell-Garnett model \cite{Choy} in its generalized form outlined by Bruggeman \cite{Bruggeman}. In this model, the permittivity of an effective medium  $\varepsilon_{eff}$ consisting of several components can be found as a solution of the equation 

\begin{equation}
\label{1}
			{\sum^N_{(i=1)}} {f_i} \theta_i (\varepsilon_{eff} - \varepsilon_{i}) = 0 ,
\end{equation}			
where $\varepsilon_{i}$ is the dielectric permittivity of the components, ${f_i}$ is the filling factor of the  $i$ component, and $\theta_i$ is the form factor of the $i$ component. The normalization condition 

\begin{equation}
\label{2}
			{\sum^N_{(i=1)}} {f_i} = 1 , {\sum^N_{(i=1)}} \theta_i = 1 
\end{equation}	
must also be met. 

According to Eq.~\eqref{1}, the anisotropy of the sample can be related both to the anisotropy of the dielectric permittivity of the microcrystals themselves and to the anisotropy of their distribution. For randomly-oriented microcrystals, the anisotropy of the sample can be associated only with the dependence of the filling factor on the direction (Eq.~\eqref{1}). Knowing the thickness of the sample and the spectral position of the isotropic points, we can determine the difference in refractive indices for ordinary and extraordinary beams.

\begin{equation}
\label{3}
\langle 2 n \rangle ({n^\perp}_{eff} - {n^\parallel}_{eff})=({{n^\perp}_{eff}})^2	- ({{n^\parallel}_{eff}})^2 = {\varepsilon^\perp}_{eff} - {\varepsilon^\parallel}_{eff}			 .
\end{equation}

From the obtained data, we can estimate the value of the anisotropy of the filling factor. The average thickness of the sample is estimated from the amount of deposited material to be 5~microns. Averaged in all directions the refractive index is taken as $\langle n \rangle \approx 2$. The presence of an optical axis means that ordinary and extraordinary rays propagate at different phase velocities. This is manifested in the conversion of linear polarization into circular polarization. If the thickness of the sample is such that the phase difference of these rays is $N\pi$, then there is no polarization conversion. This is the isotropic point in the transmission spectrum. If the phase difference is $( \pi/2 + N\pi) $, then the conversion is maximal. Some variation in the spectral position of the isotropic point in Figure~\ref{fig:CircPol} occurs as the incident light beam spot shifted slightly over the surface area when the sample was rotated.

Hence, we found the difference in the effective refractive indexes to be ${\varepsilon^\perp}_{eff} - {\varepsilon^\parallel}_{eff} \approx 0.20$. Assuming that the microcrystals are oriented completely chaotically, and the observed anisotropy is caused only by their spatial alignment, we obtain a filling factor of ${f^\perp} - {f^\parallel} \approx 0.05$ . We emphasize that this is only an estimated value based on the outlined assumptions. Such a small value may not be noticeable "by eye", but it is manifested in the transmission spectra.

We performed a statistical analysis of the surface image as described in \cite{Jacucci}. 

Figure~\ref{fig:SEM}~a clearly shows that there are light and dark spots on the surface of the sample. We believe that the darker areas are associated with voids between microcrystals.

\begin{figure}[h]
 \centering
  \includegraphics[width=0.7\linewidth]{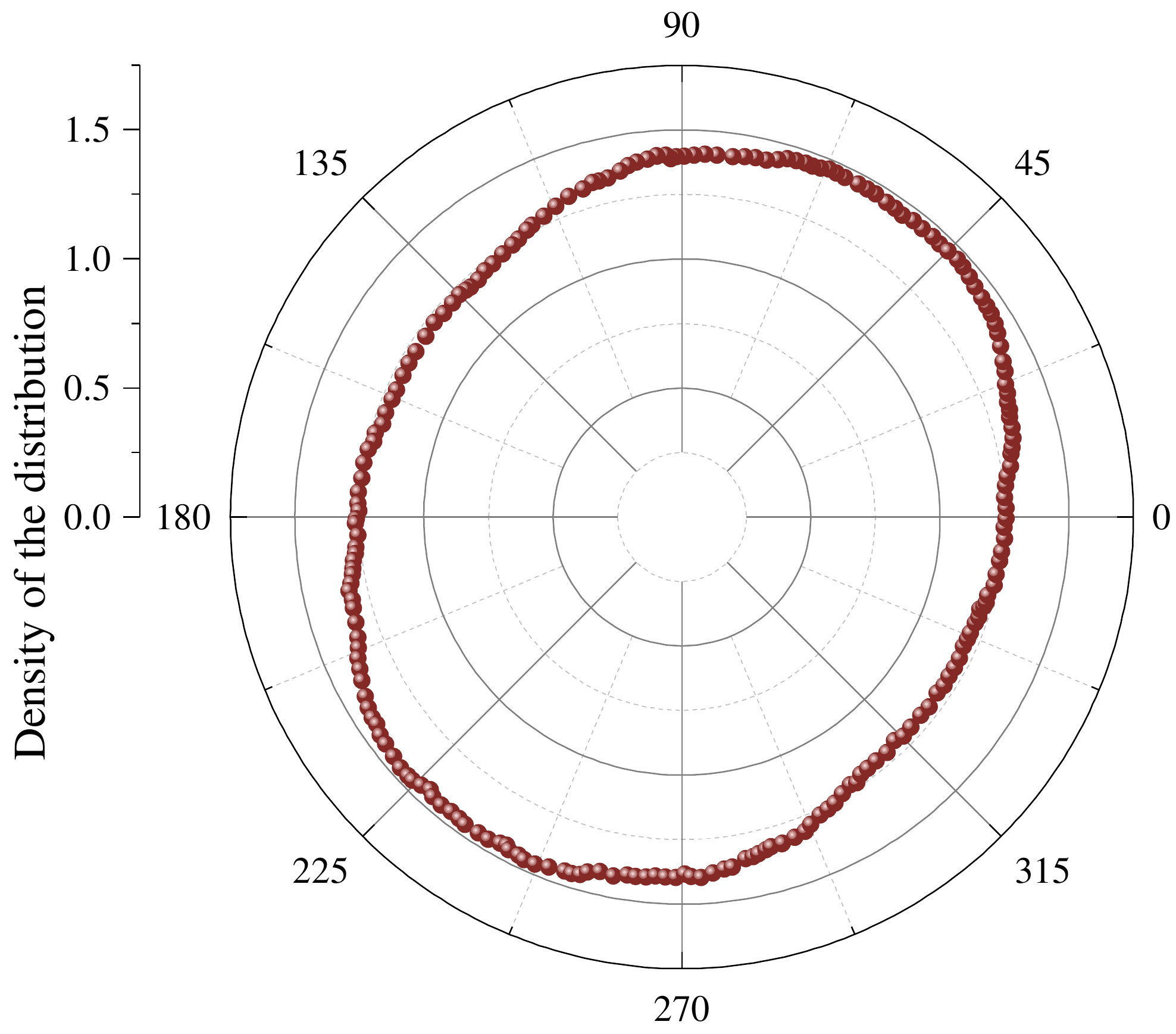}
  \caption{The relative density of the distribution of microcrystals, constructed by statistical analysis of the distribution of dark and light spots in Figure~\ref{fig:SEM}~a.
}
  \label{fig:Calcul}
\end{figure}

The spatial dependence of the distribution of dark and light spots was analyzed. The density of dark and light spots as a function of the azimuthal angle is shown in polar coordinates in Figure~\ref{fig:Calcul}.

The distribution density of microcrystals is anisotropic, with an axis directed at an angle of $30^\circ$ relative to the edge of the sample. The direction of the anisotropy axis determined coincides with the direction of the axis determined from the polarization measurements. 

The conclusion about the presence of a selected direction in the sample was confirmed using an atomic force microscope (AFM). Large-scale anisotropy of the microcrystal distribution density is clearly visible in AFM images of the surface. 

A possible reason for the macroscopic ordering is the electrostatics of the exfoliated microscrystals. When we exfoliate the bulk \textit{h}-BN crystals in the ultrasonic bath, the resulting microcrystals have an electrostatic difference between the edges where intralayer covalent bonds are broken and the planar surfaces where only weak van der Waals bonding is broken. 
They could then agglomerate in an ordered manner in accordance with the direction of their dipole momentum during filtering. Due to such anisotropic correlations in the distribution of microcrystals, their density in different directions may be different. This may be invisible to the eye (see Figure 1a in \cite{Jacucci}), but it may manifest in the polarization properties. Indeed, during the electrification of dielectric crystals, electrostatic fields may be sufficient for their ordering at large distances.

\section{Conclusions}

The transmission and reflection spectra of thin films formed of individual stacked \textit{h}-BN microcrystals were studied in the transparency range below the edge of the fundamental absorption band. The spectral dependences of the linear and circular polarization of light passed through the sample are measured. Optical and scanning electron microscopy showed the absence of noticeable selected directions in the sample. However, birefringence was observed in the study of the Stokes parameters of the light passed through the sample. It was established that the optical axes were unidirectional over the entire sample area. Simultaneously, the diffused light from the sample was not polarized. This indicates that the cause of anisotropy was not in the properties of individual microcrystals but is common for the entire sample. Statistical analysis of the optical images confirms the thin films demonstrating macroscopic ordering of the density distributions in the same direction as the optical axis.


\vspace{6pt} 



\authorcontributions{Experiments, L.K., L.A., M.Zh.; sample fabrication, B.H., A.B.; writing—original version preparation, L.K.; review, M.Zh., B.H.; supervising, A.B.; supervising, review and editing, V.K.}

\funding{This work has been supported by the Engineering and Physical Sciences Research Council (EPSRC) of the United Kingdom via the EPSRC Centre for Doctoral Training in Electromagnetic Metamaterials (Grant No. EP/L015331/1) and via Grant No. EP/N035569/1. 
LVK thanks the Russian Science Foundation (project No. 21-12-00304). 
}

\institutionalreview{Not applicable.}

\informedconsent{Not applicable.}

\dataavailability{The data that supports the findings of this study are available within the article.}

\acknowledgments{The authors are grateful to Sergey Pavlov for SEM measurements and Bogdan Borodin for AFM.}

\conflictsofinterest{The authors declare no conflict of interest.}

\end{paracol}
\reftitle{References}



\begin{thebibliography}{999}
\bibitem[Author2(year)]{Geim}
Geim, A. ; Grigorieva, I. {Van der Waals heterostructures.} {\em Nature} {\bf 2013}, {\em 499}, 419-425.
\bibitem[Author1(year)]{Moon}
Moon, P.; Koshino, M. {Electronic properties of graphene/hexagonal-boron-nitride moir\'e superlattice.} {\em Phys. Rev. B} {\bf 2014}, {\em 90}, 155406. 
\bibitem[Author2(year)]{Ronning}
Ronning,C. ;Banks,A. D.; McCarson,B. L.; Schlesser,R. ; Sitar,Z.  ; Davis,R. F.  ; Ward,B. L.  ; Nemanich,R. J.{Structural and electronic properties of boron nitride thin films containing silicon.} {\em J. Appl. Phys.} {\bf 1998}, {\em 84}, 5046.
\bibitem[Author1(year)]{Hogan} 
Hogan, B. T. ;  Kovalska, E. ;  Zhukova, M. O.;  Yildirim, M. ;  Baranov, A. ;  Craciun, M. F. ;  Baldycheva, A. {2D $WS_2$ liquid crystals: tunable functionality enabling diverse applications.} {\em Nanoscale} {\bf 2019}, {\em 11}, 16886.
\bibitem[Author2(year)]{Akbari}
Akbari, A. ; Sheath, P.; Martin, S.T.; Shinde, D.B.; Shaibani, M.; Banerjee, P.C.; Tkacz, R.; Bhattacharyya, D.; Majumder, M. {Large-area graphene-based nanofiltration membranes by shear alignment of discotic nematic liquid crystals of graphene oxide.} {\em Nat Commun.} {\bf 2016}, {\em 7}, 10891.
\bibitem[Author2(year)]{Fu}
Fu, K.; Wang, Y.; Yan, C.; Yao, Y.; Chen, Y.; Dai, J.; Lacey, S.; Wang, Y.; Wan, J.; Li, T.; Wang, Z.; Xu, Y.; Hu, L. Graphene Oxide-Based Electrode Inks for 3D-Printed Lithium-Ion Batteries. {\em J. Adv Mater.} {\bf 2016}, {\em 28}, 2587-94.
\bibitem[Author1(year)]{Jalili} 
Jalili, R.  ; Aminorroaya-Yamini, S.  ; Benedetti, T. M.  ; Aboutalebi, S. H. ;  Chao, Yu.  ; Wallace, G. G.  ; Officer, D. L. {Processable 2D materials beyond graphene: $MoS_2$ liquid crystals and fibres.} {\em Nanoscale} {\bf 2016}, {\em 8}, 16862. 
\bibitem[Author2(year)]{Shaban}
Shaban,P. ; Oparin,E. ; Zhukova,M. ; Hogan,B. ; Kovalska,E. ; Baldycheva,A. ; Tsypkin, A.{Transmission properties of van der Waals materials for terahertz time-domain spectroscopy applications.} {\em AIP Conference Proceedings} {\bf 2020}, {\em 2300}, 020111.
\bibitem[Author1(year)]{Baldycheva}
Hogan, B. T. ; Kovalska, E. ; Craciun, M. F. ; Baldycheva, A. {2D material liquid crystals for optoelectronics and photonics.} {\em J. Mater. Chem. } {\bf 2017}, {\em 5}, 11185.
\bibitem[Author1(year)]{Shin}
Lee, J. H. ; Shin, D. W. ; Makotchenko, V. G. ; Nazarov, A. S. ; Fedorov, V. E. ; Kim, Yu. H. ; Choi, J.-Y. ; Kim, J. M. ; Yoo, J.-B. One-step exfoliation synthesis of easily soluble graphite and transparent conducting graphene sheets.  {\em Adv. Mater.} {\bf 2018}, {\em 30}, 1802953. 
\bibitem[Author2(year)]{Zhukova}
 Zhukova,M.O.;  Hogan,B.T.;  Oparin,E.N.;  Shaban,P.S.;  Grachev,Y.V. ;  Kovalska,E. ;  Walsh,K.K. ;  Craciun,M.F. ; Baldycheva, A. ;  Tcypkin, A.N.{Transmission Properties of $FeCl_3$-Intercalated Graphene and $WS_2$ Thin Films for Terahertz Time-Domain Spectroscopy Applications.} {\em Nanoscale Res. Lett.} {\bf 2019}, {\em 14}, 225.
\bibitem[Author1(year)]{Grinyaev}
 Grinyaev,S. N.;  Konusov,F. V. ;  Lopatin, V.V. {Deep levels of nitrogen vacancy complexes in graphite-like boron nitride..} {\em Phys. Solid State} {\bf 2002}, {\em 44}, 286-293.
\bibitem[Author1(year)]{Ngeveniya}
Ngwenya, T. B. ; Ukpong, A. M. ; Chetty, N. {Defect states of complexes involving a vacancy on the boron site in boronitrene.} {\em Phys. Rev. B} {\bf 2011}, {\em 84}, 245425. 
\bibitem[Author2(year)]{Jacucci}
Jacucci, G.; Bertolotti, J.; Vignolini, S.{Role of Anisotropy and Refractive Index in Scattering and Whiteness Optimization.} {\em Adv. Optical Mater.} {\bf 2019}, {\em 7}, 1900980.
\bibitem[Author1(year)]{Segure}
Segura, A. ; Art\'us, L. ; Cusc\'o, R. ; Taniguchi, T. ; Cassabois, G. ; Gil, B. {Natural optical anisotropy of h-BN: Highest giant birefringence in a bulk crystal through the mid-infrared to ultraviolet range.} {\em Phys. Rev. Materials} {\bf 2018}, {\em 2}, 024001.
\bibitem[Author1(year)]{Kotova}
Kotova,L.V. ;  Platonov,A.V. ;  Poshakinsky,A.V. ;  Shubina, T.V.{Polarization Conversion in $MoS_2$ Flakes.} {\em Semiconductors} {\bf 2020}, {\em 54}, 1509. 
\bibitem[Author2(year)]{Thouless}
 Thouless, D.J.{Electrons in disordered systems and the theory of localization.} {\em Phys. Rep.} {\bf 1974}, {\em 13}, 94-142.
\bibitem[Author1(year)]{Landauer}
Landauer, R. {Electrical conductivity in inhomogeneous media.} {\em AIP Conference Proceedings} {\bf 1978}, {\em 40}, 2-45.
\bibitem[Author2(year)]{Choy}
Tuck C. Choy. \textit{Effective Medium Theory: Principles and Applications}; Publisher: Oxford: Clarendon Press.,ISBN 978-0-19-851892-1, 2016.
\bibitem[Author2(year)]{Bruggeman}
Bruggeman, D.A.G.{Berechnung verschiedener physikalischer Konstanten von heterogenen Substanzen. I. Dielektrizitätskonstanten und Leitfähigkeiten der Mischkörper aus isotropen Substanzen.} {\em J. Ann. Phys.} {\bf 1935}, {\em 416}, 636-664.







\end{thebibliography}


\end{document}